\documentclass[10pt]{article}
\usepackage[OE]{express}
\usepackage{siunitx}             
\usepackage{graphicx}            
\usepackage{xcolor}   
\usepackage{amsmath}
\usepackage[utf8]{inputenc}
\usepackage[T1]{fontenc}
\usepackage{booktabs}
\usepackage{float}
\usepackage{hyperref}  %
\hypersetup{colorlinks=true, citecolor=black, colorlinks, linkcolor=black, urlcolor=black}     
\usepackage{subfigure}

\begin{document}
\title{Atomic frequency reference at 1033\,nm for ytterbium\,(Yb)-doped fiber lasers and applications exploiting a rubidium\,(Rb) 5$S_{1/2}$ to 4$D_{5/2}$ one-colour two-photon transition}
\author{Ritayan Roy,\authormark{1,2,*} Paul C. Condylis,\authormark{1} Yik Jinen Johnathan,\authormark{1,3} and Bj\"orn Hessmo\authormark{1,4}}

\address{\authormark{1}Centre for Quantum Technologies (CQT), 3 Science Drive 2, Singapore 117543\\
\authormark{2}Present address: School of Physics and Astronomy, University of Southampton, Highfield, Southampton, SO17 1BJ, United Kingdom\\

\authormark{3}Present address: Institute of High Performance Computing, 1 Fusionopolis Way, \#16-16 Connexis, Singapore 138632\\
\authormark{4}Department of Physics, National University of Singapore, 2 Science Drive 3, Singapore 117542}

\email{\authormark{*}ritayan.roy@u.nus.edu} 

\date{\today}

\begin{abstract}
We demonstrate a two-photon transition of rubidium\,(Rb) atoms from the ground state\,(5$S_{1/2}$) to the excited state\,(4$D_{5/2}$), using a home-built ytterbium\,(Yb)-doped fiber amplifier at 1033\,nm. This is the first demonstration of an atomic frequency reference at 1033\,nm as well as of a one-colour two-photon transition for the above energy levels. A simple optical setup is presented for the two-photon transition fluorescence spectroscopy, which is useful for frequency stabilization for a broad class of lasers. This spectroscopy has potential applications in the fiber laser industry as a frequency reference, particularly for the Yb-doped fiber lasers. This two-photon transition also has applications in atomic physics as a background-\,free high-\,resolution atom detection and for quantum communication, which is outlined in this article. 
\end{abstract}




 \newcommand{\noop}[1]{}

\section{Introduction}
Fiber lasers operating in the range of 1000 to 1200\,nm offer both very good energy efficiency as well as the ability to generate high powers.  Additionally, such lasers may also be created with extremely narrow line-widths. These sources are already in use for several applications, including photochemistry, medicine, and communications. At this wavelength band, 1000 to 1200\,nm, there are few atomic or molecular frequency references available for wavelength stabilization~\cite{wallerand_2006}.

We demonstrate the use of rubidium\,(Rb) as a frequency reference at 1033\,nm, exploiting the 5$S_{1/2}$ to 4$D_{5/2}$ two-photon transition. This transition can be driven by Yb-doped fiber lasers or amplifiers, and appears to be a suitable atomic frequency reference for such light sources. 

In the field of atomic physics, a fiber laser of 1033\,nm wavelength off resonant to the Rb 5$S_{1/2}$ to 4$D_{5/2}$ two-photon transition could be useful for constructing optical lattices and dipole traps, due to its high power and frequency stability. The on resonant beam to the above two-photon transition would be useful for highly localized and non-linear excitation of the atoms. This would allow the high-resolution imaging of Rb atoms to be carried out. It would also be useful for the background free detection of atoms~\cite{Ohadi_2009} using single colour or multi-colour two-photon transition or by ladder excitations. 

This two-photon transition has applications in precision spectroscopy as it is possible to eliminate the Doppler broadening\,(first-order), which is around one hundred to one thousand times the natural linewidth~\cite{Vasilenko_1970, Cagnac_1973}. 

Finally, this transition also has applications in the fields of quantum information and quantum communications ~\cite{Chaneliere_2006}, which we describe in a later section of this article.

\section{Two-photon transition probability for rubidium}
For a two-photon transition, the atom from the ground state\,($g$) is excited to the excited state\,($e$) by absorbing simultaneously two photons with equal angular frequencies $\omega$, where, $(E_{e}-E_{g})=2\hbar\omega$. During the two-photon excitation the real intermediate state\,($r$) is not populated. The first two-photon absorption in the optical domain was observed by Abela~\cite{Abella_1962} in cesium\,(Cs) vapour. Using a thermally tuned ruby laser, by absorption of two photons at 653.55 nm, the transition between the 6$S_{1/2}$ ground state to the 9$D_{3/2}$ excited state of Cs was investigated. The detection scheme was quite straightforward. The photon spontaneously emitted from the excited state\,($e$), decays via the real intermediate state\,($r$). The spontaneously emitted photons have different wavelengths than the excitation photons. By collecting those spontaneously emitted photons, the two-photon transition spectroscopy is observed. 

\begin{figure}[tb]
    \centerline{\includegraphics[width=0.7\textwidth]{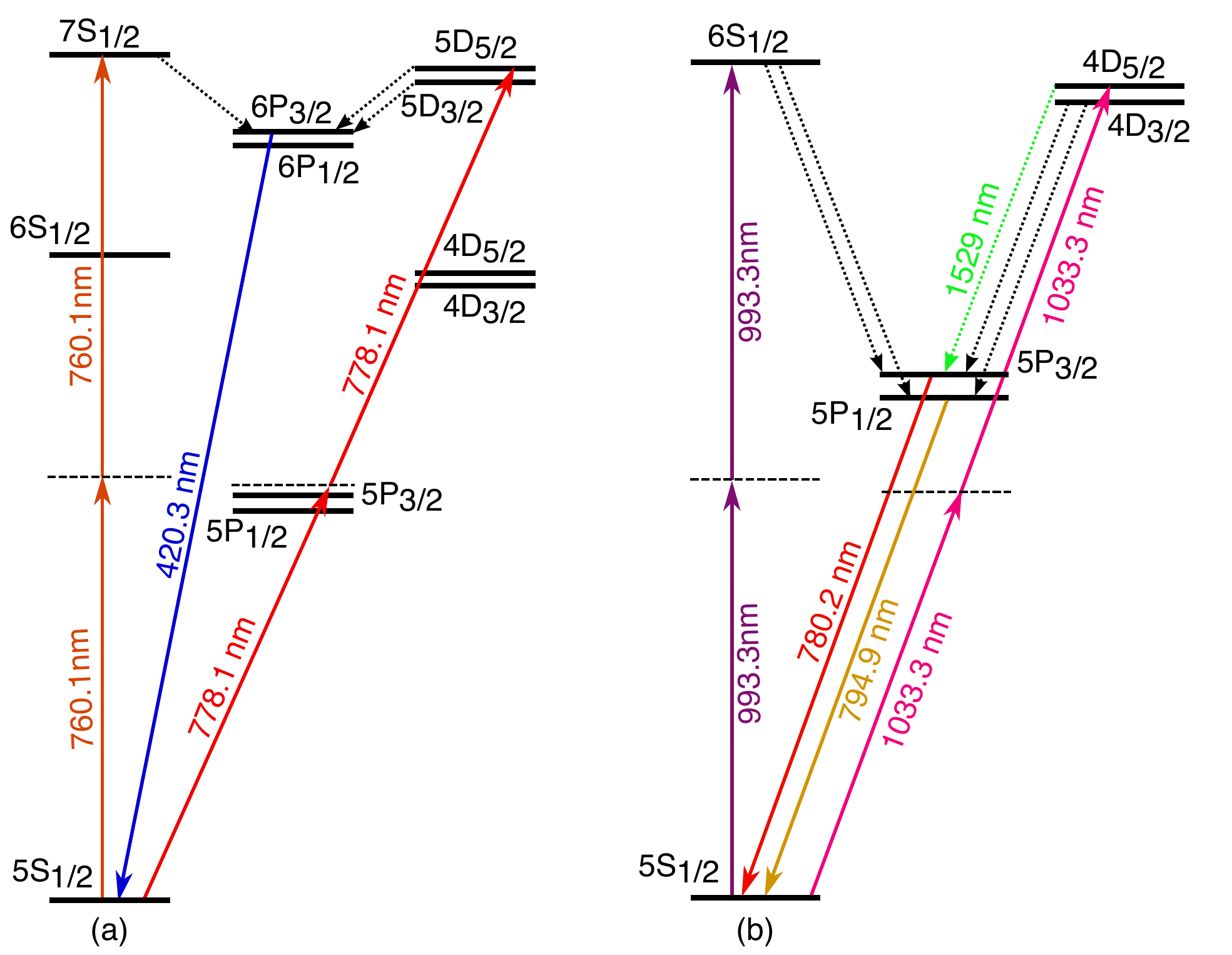}}
   \caption{ \label{fig:All-TP-EnergyLevelNew}(a) Rb 5$S$ to 5$D$ and 5$S$ to 7$S$ two-photon transition schemes. For both these transitions, the respective virtual intermediate levels\,(shown in dotted line), is blue-detuned from the 5$P_{3/2}$ state and the atom decays back to the ground state from the excited state via 6$P_{3/2}$ level emitting photon of 420.3\,nm. (b) Rb 5$S$ to 4$D$ and 5$S$ to 6$S$ two-photon transition schemes. For both these transitions, the respective virtual intermediate levels\,(shown in dotted line), is red-detuned from the 5$P_{3/2}$ state and the atom decays back to the ground state from the excited state via 5$P_{3/2}$ level emitting photon of 780.2\,nm. For both the Figs. 5$P_{3/2}$ state represents the real intermediate state\, ($r$). }
\end{figure}

We categorise the two-photon transitions of rubidium\,(Rb) with respect to the 5$P_{3/2}$\,(intermediate) level as follows:

\noindent
\begin{enumerate}
\item The transitions in which the virtual intermediate level\,(dashed lines) is blue-detuned from the 5$S_{1/2}$ to 5$P_{3/2}$ transition:
The two-photon transitions from the 5$S$ to 5$D$ and 5$S$ to 7$S$, have their respective virtual intermediate states, blue-detuned from the 5$S_{1/2}$ to 5$P_{3/2}$ transition as illustrated in the Fig.~\ref{fig:All-TP-EnergyLevelNew}(a). We shall refer these two-photon transitions as \textit{blue-detuned two-photon transitions}; and,

\item The virtual intermediate level is red-detuned from the 5$S_{1/2}$  to 5$P_{3/2}$ transition:
The two-photon transitions from 5$S$ to 4$D$ and 5$S$ to 6$S$, have their respective virtual intermediate states, red-detuned from the 5$S_{1/2}$ to 5$P_{3/2}$ transitions as illustrated in the Fig.~\ref{fig:All-TP-EnergyLevelNew}(b). We shall refer these two-photon transitions as \textit{red-detuned two-photon transitions} through out this article.

\end{enumerate}

The Doppler-free\,(1st order) two-photon transition probability is calculated using a second order perturbation theory by G. Grynberg and B. Cagnac~\cite{Grynberg_1977}.
Using their method the two-photon transition probabilities for the blue-detuned and the red-detuned two-photon transitions for Rb atoms, are calculated and tabulated in Table~\ref{table:TrnsProb}. 
\begin{table}[h]
\centering
\caption{The two-photon transition probabilities for various excited states of Rb. The transition probability is normalized to the 5S to 5D transition probability.}
\label{table:TrnsProb}
\resizebox{1.0\textwidth}{!}{ 
\begin{tabular}{|c|c|c|c|c|c|c|}
\hline
\begin{tabular}[c]{@{}c@{}}Two-photon\\ transition \\ levels\,(Rb)\end{tabular} & \begin{tabular}[c]{@{}c@{}}Wave-\\length\\ (nm)\end{tabular} & \begin{tabular}[c]{@{}c@{}}Beam \\ power\\ (mW)\end{tabular} & \begin{tabular}[c]{@{}c@{}}Beam \\ diameter \\ ($\mu$m)\end{tabular} & \begin{tabular}[c]{@{}c@{}}Energy \\ defect $\Delta\omega_{r}$\\(THz)\end{tabular} & \begin{tabular}[c]{@{}c@{}}Transition\\ probability\end{tabular} & \begin{tabular}[c]{@{}c@{}}Normalized\\ transition\\ probability\end{tabular} \\ \hline
5S to 5D & 778.1 & 10 & 30 & 6.6 & 2.2$\times10^{4}$ & 1 \\ \hline
5S to 7S & 760.1 & 10 & 30 & 64 & 220 & 1/100 \\ \hline
5S to 6S & 993.3 & 10 & 30 & -470 & 8 & 1/3000 \\ \hline
5S to 4D & 1033.3 & 10 & 30 & -550 & 7 & 1/3400 \\ \hline
5S to 4D & 1033.3 & 100 & 30 & -550 & 650 & 1/34 \\ \hline
\end{tabular}
}
\end{table}
\\

\section{Experimental Setup} 
\begin{figure}[b]
    \centerline{\includegraphics[width=0.7\textwidth]{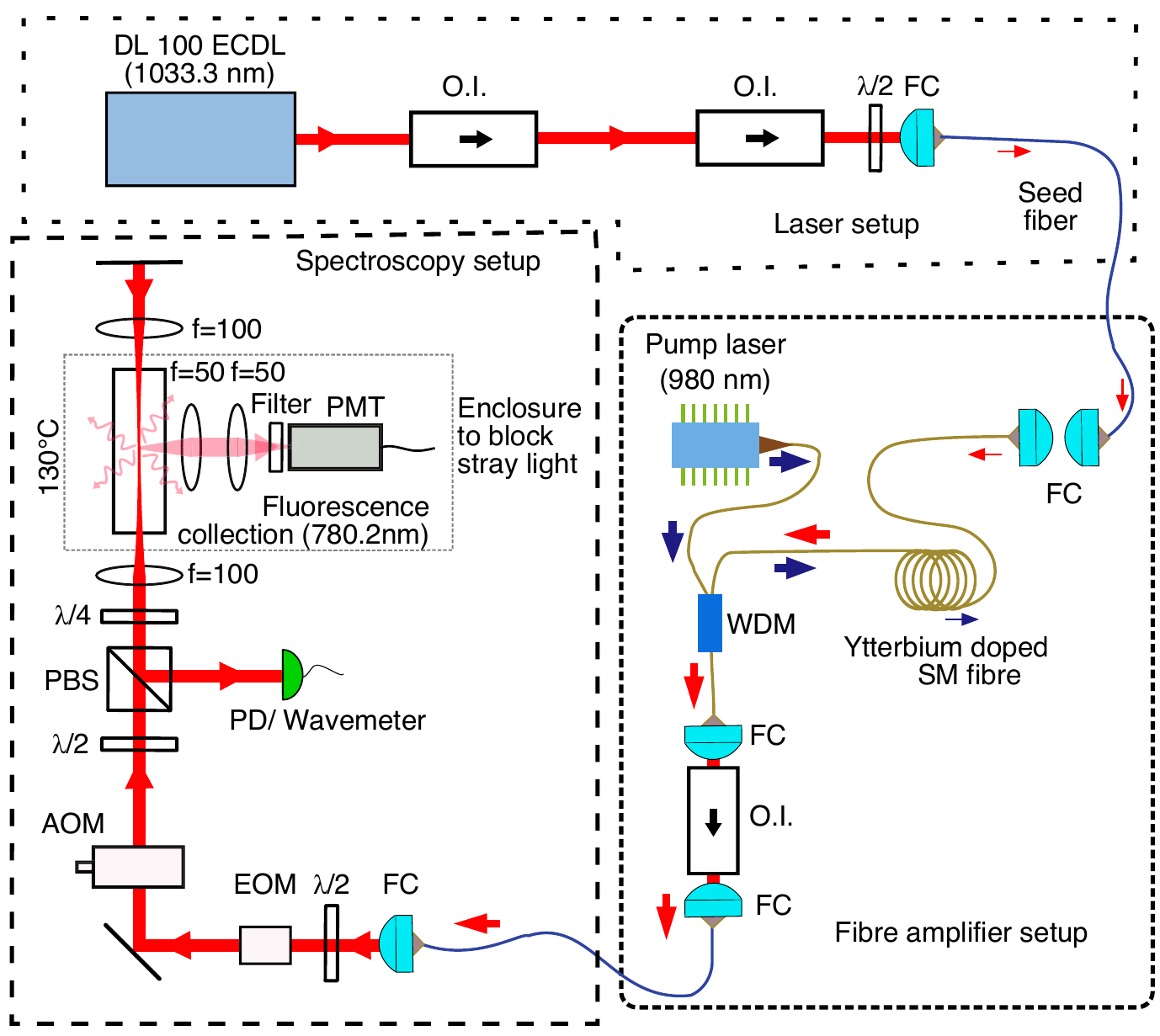}}
    \caption{\label{fig:TwoPhotonFull} At the top, the layout of the seed laser setup. At the bottom right, the layout of the fiber amplifier is shown. At the bottom left the spectroscopy setup is shown. PD: photodetector; $\lambda$/n: $\lambda$/n wave plate; f: focal length of lens in mm; OI: optical isolator; PBS: polarization beam splitter; FC: fiber coupler, FP: Fabry-Perot, WDM: wavelength-division multiplexer, AOM: acusto-optical modulator; EOM: electro-optical modulator; PMT: photomultiplier tube, Filter: 780\,nm interference filter. Thick arrows signify higher power.}
\end{figure}

A home built fiber laser amplifier is used to excite atoms from both 5$S_{1/2}$ to 4$D_{5/2}$ and 5$S_{1/2}$ to 4$D_{3/2}$ transitions using a ytterbium-doped silica fiber. Ytterbium-doped silica fiber has a very broad absorption spectrum from 800\,nm to 1064\,nm and emission bands from 970\,nm to 1200\,nm. It offers a very efficient operation with narrow linewidth and convenient means of amplification, as an amplifier, of a particular wavelength in the emission band assisted with a wide variety of pump lasers.\\
A Toptica DL 100 ECDL is used as a seed laser, with tuning wavelength ranging from 980 to 1075\,nm. The grating is tuned to the wavelength at 1033.3\,nm. The output power of the seed laser is around 50\,mW with a typical mode-hop free tuning range of around 10\,GHz. Current and temperature tuning of the diode laser provides further control of the output wavelength.
 The shape of the beam emitted from the seed laser diode is modified from elliptical to circular using a cylindrical lens pair. The beam then passes through two optical isolators, each providing 35\,dB isolation. Then, the beam is coupled to a polarization-maintaining\,(PM) optical fiber to seed the fiber amplifier. It is essential to have both the optical isolators, in order to prevent any optical feedback to the laser diode coming from the fiber amplifier, in this case via the seed fiber as shown in Fig.~\ref{fig:TwoPhotonFull}. After leaving the first optical isolator a small part of the beam is diverted in a home built Fabry-Perot\,(FP) cavity of finesse 50, where the single-frequency operation of the laser is monitored, not shown.

\subsection{Fiber amplifier}
A ytterbium-doped silica fiber is used in order to amplify the laser power. The fiber is a core pumped, non-PM, and single mode (Liekki YB1200-4/125). It can generate a moderate power of 250\,mW, which is sufficient for our spectroscopy power requirements. Ytterbium-doped silica fiber is a good candidate~\cite{ZhiyiWei_2012} for amplifying lasers in the range of 1000-1100\,nm with ours being at 1033.3\,nm. The fiber amplifier is seeded with 15\,mW power through a optical fiber, called the \textit{seed fiber}. The ytterbium-doped silica fiber is pumped with a strong 400\,mW pump beam of 980\,nm\,(3S Photonics, 1999HPM). This amplification process is done using a wavelength-division multiplexer\,(WDM) by mixing the counter-propagation pump and seed beams as shown in the Fig.~\ref{fig:TwoPhotonFull} fiber amplifier setup. The pump laser\,(980 nm), marked with blue arrow, get absorbed in the ytterbium-doped fiber. The seed light\,(1033.3\,nm), marked with the red arrows, enters the ytterbium-doped fibre from the opposite end of the pump beam. The seed power of 15\,mW is amplified to around 250\,mW, measured at the output of the WDM. This amplified beam is then sent to the spectroscopy setup via an optical isolator and a PM optical fibre.

\subsection{Fluorescence spectroscopy}

In the spectroscopy setup frequency modulation is applied to the light via an Electro-Optic-Modulator\,(EOM), Photonics EOM-01-12.5-V, with a resonance frequency of 12.5\,MHz. By sending sufficient radio-frequency\,(rf) power through the EOM, detectable sidebands are generated around the main signal. The sidebands are used as a frequency marker, to measure the hyperfine splitting\,(HFS) in the excited state as discussed in Section~\ref{SpectroscopyResult}. An AOM,\,(AA Opto-Electronic MT110-A1-IR), +1 order, is driven by 110\,MHz rf source to deflect the laser beam, shifting the frequency of the light by 2$\times$110\,MHz for the retro-reflected beam. This prevents unwanted feedback in to the fiber amplifier~\cite{Ryan_1993}. 

The $\sigma^{+}$ and $\sigma^{+}$ polarization configuration is used for the retro-reflected beams of the spectroscopy to maximise the signal strength~\cite{Olson_2006}. The beam is turned into $\sigma^{+}$ , by using a $\lambda$/2 WP, a PBS, and a $\lambda$/4 WP. A laser power of around 140\,mW enters in the rubidium cell. The beam is focused on a narrow spot\,(diameter 30-50\,$\mu$m) inside the cell using a pair of collimating lenses of focal length 100\,mm. These focal lengths are selected to fit the beam around the dimension of the Rb cell. Reflection from a mirror provides the required counter-propagating laser beam. The two beams are carefully aligned to overlap inside the Rb cell. 

The Rb spectroscopy cell is kept at a constant temperature of approximately 130$^{\circ}$C by wrapping a heating tape around it. Fine tuning of the laser wavelength is done to match the two-photon transition to generate the fluorescence signal. The signal is then collected onto a photo-multiplier tube\,(PMT), Hamamatsu H10722-01, using a pair of 50\,mm focal length lenses. A 780\,nm interference filter is used to reduce interference from scattered light. One part of the collected signal on photomultiplier tube is used to generate an error signal for locking the laser. The other part is used to record the signal for analysis. The setup is used to detect near infrared fluorescence at 780.2\,nm from the 4$D_{5/2}$ to 5$S_{1/2}$ or the 4$D_{3/2}$ to 5$S_{1/2}$ transition via 5$P_{3/2}$ cascade decay as shown in Fig.~\ref{fig:All-TP-EnergyLevelNew}(b). 
\begin{figure}[t]
    \centerline{\includegraphics[width=0.6\textwidth]{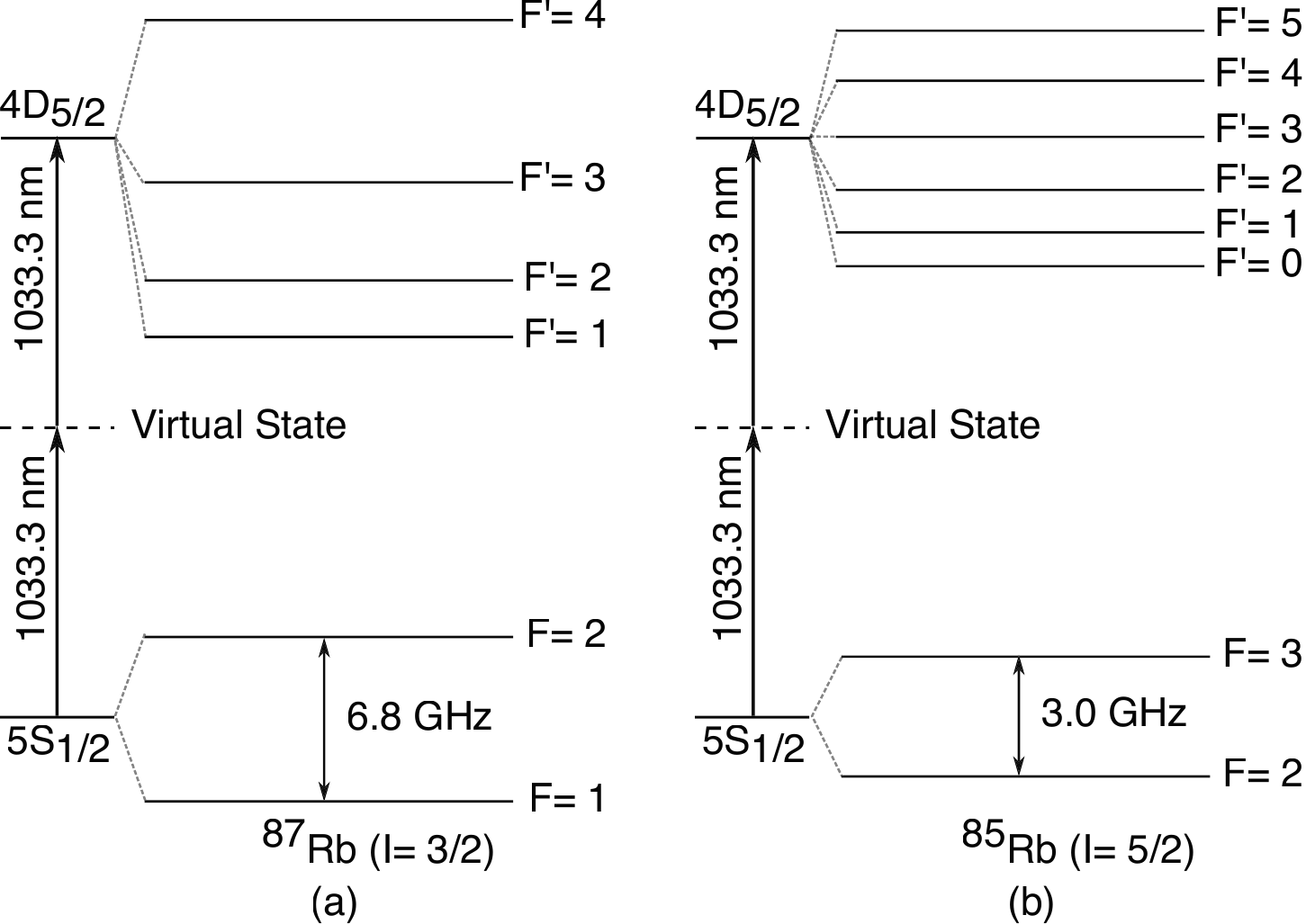}}
    \caption{\label{fig:Hyperfine} (a) The hyperfine splitting for the $^{87}$Rb, 5$S_{1/2}$ to 4$D_{5/2}$ transitions and (b) for the $^{85}$Rb, 5$S_{1/2}$ to 4$D_{5/2}$ transitions. For the two-photon transition the allowed transitions are ${\Delta F=0, \pm 1, \pm 2}$.}
\end{figure}
As represented by the dotted line shown in Fig.~\ref{fig:TwoPhotonFull}, the Rb cell is surrounded by a metal enclosure. Two small holes provide access and exits for the incident and retro-reflected laser beams. This helps to block stray light from the room thereby reducing the background noise. A part of the retro-reflected beam, reflected by the PBS, is collected on a photodetector and a optical wavemeter for monitoring power and laser wavelength respectively. 

\section{Spectroscopy Result} \label{SpectroscopyResult}

\begin{figure}[b]
    \centerline{\includegraphics[width=0.5\textwidth]{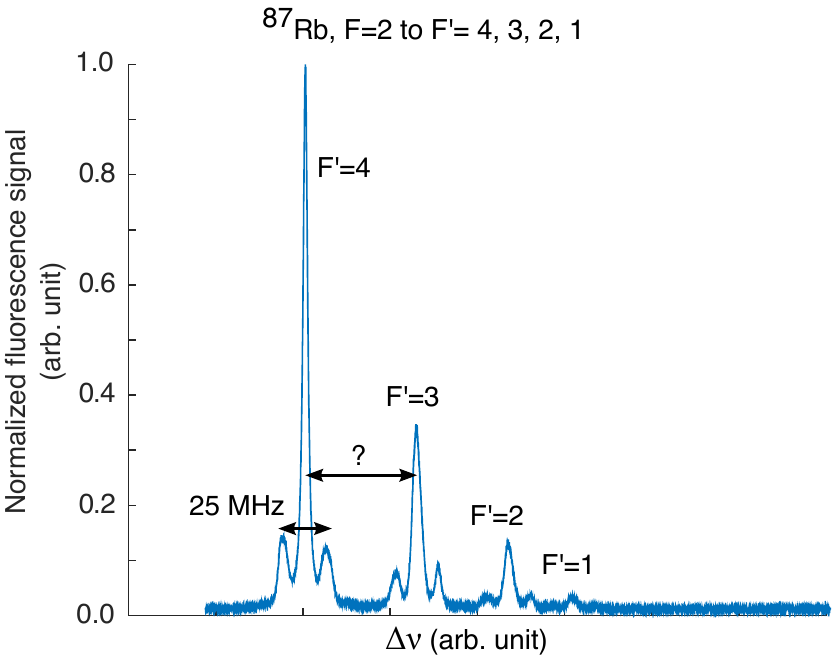}}
    \caption{\label{fig:SideBand} Sidebands of 12.5\,MHz is used as the frequency marker for the measurement of the hyperfine splittings. Around 100 spectra are used to determine the frequency scale.}
\end{figure}
The rubidium vapour cell contains both the species 85 and 87, thus both the $^{87}$Rb and $^{85}$Rb, 5$S_{1/2}$ to 4$D_{5/2}$ two-photon transitions are observed and the hyperfine splittings are measured. An extensive study of the Rb HFS is provided here ~\cite{Arimondo_1977}.

\begin{figure}[t]
    \centerline{\includegraphics[width=1\textwidth]{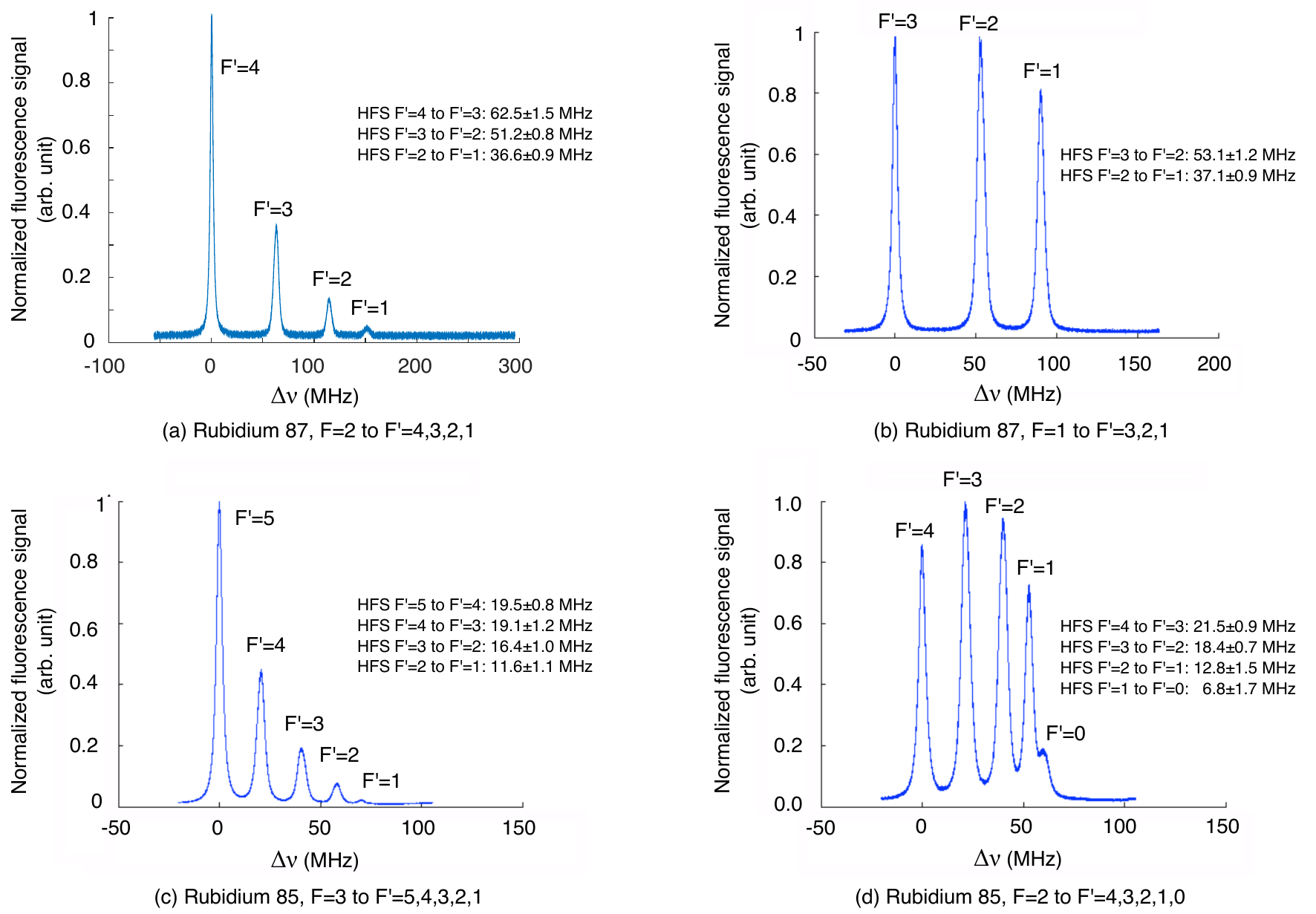}}
    \caption{\label{fig:Rbfivebytwo} The hyperfine splitting\,(HFS) of the $^{87}$Rb and $^{85}$Rb 4$D_{5/2}$ state is measured by a one-colour two-photon excitation from the (a) $^{87}$Rb ground hyperfine level F=2, (b) $^{87}$Rb ground hyperfine level F=1, (c) $^{85}$Rb ground hyperfine level F=3, and (d) $^{85}$Rb ground hyperfine level F=2. The latest reported HFS values from literature ~\cite{Lee_2015} for the $^{87}$Rb (F$^{\prime}$=4) to (F$^{\prime}$=3)= 63.826 MHz, $^{87}$Rb (F$^{\prime}$=3) to (F$^{\prime}$=2) = 52.188 MHz, $^{85}$Rb (F$^{\prime}$=5) to (F$^{\prime}$=4)= 20.749 MHz, and $^{85}$Rb (F$^{\prime}$=4) to (F$^{\prime}$=3)= 20.462 MHz.}
\end{figure}

\begin{figure}[t]
    \centerline{\includegraphics[width=0.7\textwidth]{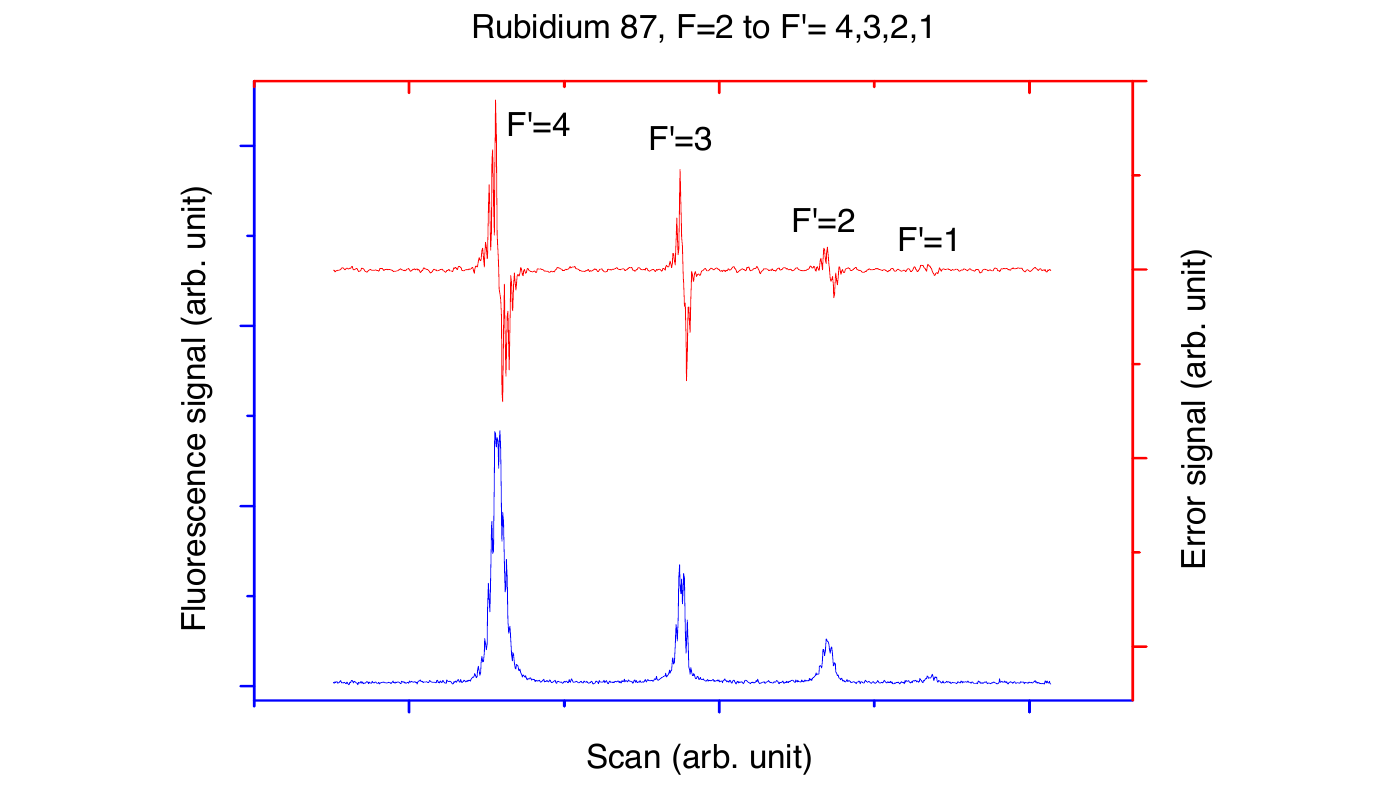}}
\caption{\label{fig:errorsignal} The error signal of the $^{87}$Rb, F=2 to F$^{\prime}$=4, 3, 2, 1 transitions without averaging. }
\end{figure}

The hyperfine energy splitting occurs from the coupling of the total electron angular momentum \textbf{J} with the total nuclear angular momentum \textbf{I}. The hyperfine interaction is represented by the Hamiltonian,
$$
\label{eq:HFS}
H_{HFS}=A\mathbf{I}\cdot \mathbf{J}+B\frac{3\left ( \mathbf{I}\cdot \mathbf{J} \right )^{2}+\frac{3}{2}\left ( \mathbf{I}\cdot \mathbf{J}\right )-I\left ( I+1 \right )J\left ( J+1 \right )}{2I\left ( 2I-1 \right )J\left ( 2J-1 \right )},
$$
where, $A$ is the magnetic dipole HFS constant, $B$ is the electric quadrupole HFS constant, $I$ and $J$ are the nuclear spin and the total electron angular momentum quantum numbers respectively. 
The hyperfine energy splittings for both $^{87}$Rb and $^{85}$Rb are shown in the Fig.~\ref{fig:Hyperfine} and in the case of two-photon transition, the allowed transitions are ${\Delta F=0, \pm 1, \pm 2}$.

By scanning the frequency of the seed laser, we record the two-photon transition signal by collecting fluorescence photons at 780.2\,nm, as shown in the Fig.~\ref{fig:All-TP-EnergyLevelNew}(b). 

Using a 12.5\,MHz EOM, sidebands are generated around the main transition signals as shown in Fig.~\ref{fig:SideBand}, e.g., for $^{87}$Rb, 5$S_{1/2}$, F=2 to 4$D_{5/2}$ hyperfine transitions. These sidebands provide a frequency reference for measuring the unknown hyperfine splittings. 
We fit the experimental data with the Lorentzian multipeak fit routines to determine the centers of spectroscopy peaks. Around 100 spectra are used for the fitting to minimize the statistical error.

Using this procedure, the hyperfine splittings are measured for both the $^{87}$Rb and $^{85}$Rb, 5$S_{1/2}$ to 4$D_{5/2}$ two-photon transitions and the findings are provided below in the Fig.~\ref{fig:Rbfivebytwo}. The error bars provided with the HFS values are coming from the fitting routine.

The error signal of the $^{87}$Rb, F=2 to F$^{\prime}$=4, 3, 2, 1 two-photon transitions are obtained by frequency modulation\,(FM) fluorescence spectroscopy as shown in Fig.~\ref{fig:errorsignal}. We have found that it is possible to lock the laser at most of the hyperfine transitions.

\section{Discussion}
The hyperfine splittings measured in this experiment are consistent with the measurements by Lee $et\ al.$~\cite{Lee_2007}, Wang $et\ al.$~\cite{Wang_2014} and recently even more precisely by Lee $et\ al.$ ~\cite{Lee_2015}. However, compared to those measurements, which require multiple laser wavelengths to execute, our technique is simpler requiring only one laser leading to a one-colour two-photon transition. Another striking difference is that we do not perform any ladder like excitations for the Rb 5$S_{1/2}$ to 4$D_{5/2}$ two-photon transition.

The higher error readings in the HFS measurements\,(Fig.~\ref{fig:Rbfivebytwo}), compared to the other measurements ~\cite{Lee_2007, Wang_2014, Lee_2015},  arise due to the fact that our assumption of linear frequency scaling with laser PZT scan voltage is not a perfect approximation. 
The Doppler-free fluorescence spectroscopy in this experiment could be improved by using magnetic shielding, using more laser power and isolating the spectroscopy system from mechanical and acoustic perturbations. 

This two-photon transition not only provides a frequency reference for Yb-doped fiber laser as well as it has various possible applications in the atomic physics experiments. We briefly describe the possible applications below.  

\subsection{High-resolution imaging and selective excitation using two-photon excitation}
\begin{figure}[htb]
  \centerline{\includegraphics[width=1\textwidth]{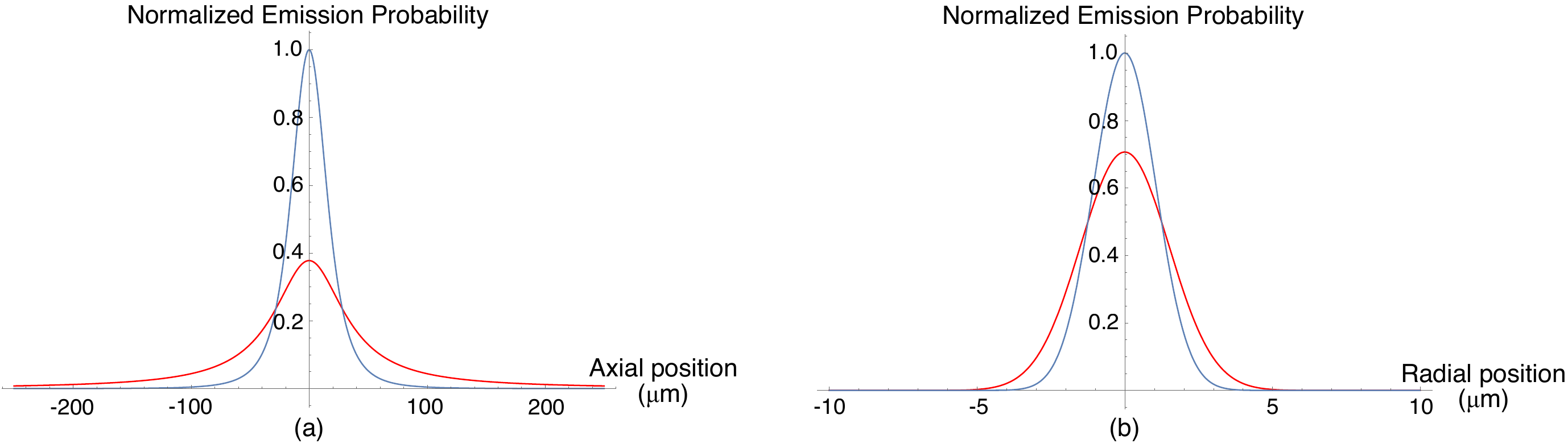}}
    \caption{\label{fig:ExcitationAxialRadial}  The two-photon excitation, using a focused dipole trap beam, where the excitation is localized to the Rayleigh volume, deactivating the fluorescence from the other part of the dipole beam. This non-linear imaging of atoms, beats the diffraction limit and provides high-resolution. The two-photon emission probability\,(blue) and single photon emission probability\,(red) along: (a) the axial and (b) the radial directions are plotted. The theoretical plots, for a 3\,$\mu$m spot size, show that using the two-photon transition scheme, which is proportional to the square of the intensity, it is possible to achieve higher resolution.}
\end{figure}

In the two-photon transition, the excitation probability is proportional to the square of the intensity as shown in the the article by G. Grynberg and B. Cagnac~\cite{Grynberg_1977}. Therefore, the fluorescence emission is localized to the Rayleigh volume of a focused dipole trap beam, while depleting fluorescence from other parts of the excitation beam as shown in Fig.~\ref{fig:ExcitationAxialRadial}.

The non-linear excitation beats the diffraction limit and provides high-resolution detection. Exploiting this phenomenon, we can also selectively excite few atoms in the cloud for probing and manipulation as the atoms will maximally interact at the focal point of the excitation beam. It is also possible to detect a single atom using this non-linear excitation and the detailed discussion is provided in the thesis~\cite{Roy_2015}.

\subsection{Creation of a dipole trap using an off-resonant laser\,( 1033\,nm) to the Rb 5S$_{1/2}$ to 4D$_{5/2}$ two-photon transition} \label{FeasibilityTwoPhotonDT}  
For a Gaussian beam propagating in free space, the spot size $w(z)$ will have a minimum value $w_{0}$ at one place along the beam axis, known as the beam waist. At a distance $z$, for a beam of wavelength $\lambda$, along the beam from the beam waist, the variation of the spot size is given by,
\begin{equation}
\label{eq:SoptSize}
w_{z}=w_{0}\sqrt{1+ \left ( \frac{z}{z_{R}} \right) ^{2}},
\end{equation}
where, $z_{R}=(\pi w_{0}^{2})/\lambda$ is called the Rayleigh range/length. 
For a focused beam, at a distance equal to the Rayleigh range\,($z_{R}$) from the waist, the width $w$ of the beam is given by $w(\pm z_{R})=\sqrt 2 w_{0}$. 
For a red-detuned dipole trap\,(w.r.t. Rb cooling transition) with a focused 1033\,nm laser, 1\,mK trap depth is achievable with a moderate power of 300\,mW\,(1200\,mW) and a beam waist of 5\,$\mu$m\,(10 $\mu$m). The scattering rate and the Stark shift is calculated to be 9\,photons/sec and 23\,MHz respectively. It would be possible to excite selectively small fraction of the atoms in a dipole trap, using two-photon excitation, without perturbing others.

\subsection{Application towards the quantum optics and communication}

Another application of our work would be in the field of quantum optics and communication. Using this two-photon transition in Rb, it would be possible to generate time correlated and entangled photon pairs, which are important for the applications in quantum information and communication~\cite{Srivathsan_2013,GulatiPRA_2014}.

The photon pairs generated from the two steps cascade decay from the 4$D_{5/2}$ to 5$P_{3/2}$ at wavelength 1.5\,$\mu$m and from 5$P_{3/2}$ to 5$S_{1/2}$ at 780\,nm should be correlated in time as shown in the Fig.~\ref{fig:All-TP-EnergyLevelNew}(b). The top photon of the cascade\,(1.5 $\mu$m) is suitable for long distance quantum communication whereas the bottom photon of the cascade\,(780\,nm) can be used for mapping long lived quantum memory~\cite{Chaneliere_2006}. As this one-colour two-photon transition is a Doppler-free process, the generated photons will therefore be narrowband. The narrow bandwidth of photons is perquisite for efficient atom-photon interaction~\cite{Srivathsan_2013,Chaneliere_2006}. 

A single laser can be employed to excite the 5$S_{1/2}$ to 4$D_{5/2}$ transition for two-photon excitation, instead of two different lasers required for the ladder excitation. All the laser wavelengths are far away from each other, so, the emitted photons would be easily separable from each other using optical filters as well as from the two-photon excitation light. This will also help to considerably reduce the number of background photons polluting the collection modes.

\section{Conclusion}
In this article an atomic frequency reference at 1033\,nm, using a rubidium two-photon transition is demonstrated. This would be useful in industry and photonics community as a frequency reference for Yb-doped fibre lasers. We have also presented a first ever demonstration of a one-colour two-photon transition of the Rb 5$S_{1/2}$ to 4$D_{5/2}$. A feasibility study of this two-photon transition in atomic physics experiments are also discussed. This finding has wide ranging potential applications in background-free high-resolution imaging of Rb atoms even single atoms, in quantum optics and communication. 

\section*{Acknowledgments}
This research has been supported by the National Research Foundation \& Ministry of Education, Singapore. The authors acknowledge useful discussions with Gurpreet Kaur Gulati and Dzmitry Matsukevich. 


\begin{thebibliography}{10}
\newcommand{\enquote}[1]{``#1''}

\bibitem{wallerand_2006}
J.-P. Wallerand, L.~Robertsson, L.-S. Ma, and M.~Zucco, \enquote{Absolute
  frequency measurement of molecular iodine lines at 514.7 nm, interrogated by
  a frequency-doubled yb-doped fibre laser,} Metrologia \textbf{43}, 294
  (2006).

\bibitem{Ohadi_2009}
H.~Ohadi, M.~Himsworth, A.~Xuereb, and T.~Freegarde, \enquote{Magneto-optical
  trapping and background-free imaging for atoms near nanostructured surfaces,}
  Opt. Express \textbf{17}, 23003--23009 (2009).

\bibitem{Vasilenko_1970}
L.~Vasilenko, V.~Chebotaev, and A.~Shishaev, \enquote{Line shape of two-photon
  absorption in a standing-wave field in a gas,} ZhETF Pisma Redaktsiiu
  \textbf{12}, 161 (1970).

\bibitem{Cagnac_1973}
B.~Cagnac, G.~Grynberg, and F.~Biraben, \enquote{Spectroscopie d'absorption
  multiphotonique sans effet doppler,} Journal de Physique \textbf{34},
  845--858 (1973).

\bibitem{Chaneliere_2006}
T.~Chaneli\`ere, D.~N. Matsukevich, S.~D. Jenkins, T.~A.~B. Kennedy, M.~S.
  Chapman, and A.~Kuzmich, \enquote{Quantum telecommunication based on atomic
  cascade transitions,} Phys. Rev. Lett. \textbf{96}, 093604 (2006).

\bibitem{Abella_1962}
I.~D. Abella, \enquote{Optical double-photon absorption in cesium vapor,} Phys.
  Rev. Lett. \textbf{9}, 453--455 (1962).

\bibitem{Grynberg_1977}
G.~Grynberg and B.~Cagnac, \enquote{Doppler-free multiphotonic spectroscopy,}
  Rep. Prog. Phys \textbf{40}, 791 (1977).

\bibitem{ZhiyiWei_2012}
Z.~Wei, B.~Zhou, C.~Xu, X.~Zhong, Y.~Zhang, Y.~Zou, and Z.~Zhang, \emph{All
  Solid-State Passively Mode-Locked Ultrafast Lasers Based on Nd, Yb, and Cr
  Doped Media} (INTECH Open Access Publisher, 2012).

\bibitem{Ryan_1993}
R.~E. Ryan, L.~A. Westling, and H.~J. Metcalf, \enquote{Two-photon spectroscopy
  in rubidium with a diode laser,} J. Opt. Soc. Am. B \textbf{10}, 1643--1648
  (1993).

\bibitem{Olson_2006}
A.~J. Olson, E.~J. Carlson, and S.~K. Mayer, \enquote{Two-photon spectroscopy
  of rubidium using a grating-feedback diode laser,} Am. J. Phys. \textbf{74}, 218--223 (2006).

\bibitem{Arimondo_1977}
E.~Arimondo, M.~Inguscio, and P.~Violino, \enquote{Experimental determinations
  of the hyperfine structure in the alkali atoms,} Rev. Mod. Phys. \textbf{49},
  31--75 (1977).

\bibitem{Lee_2007}
W.~K. Lee, H.~S. Moon, and H.~S. Suh, \enquote{Measurement of the absolute
  energy level and hyperfine structure of the $^{87}\mathrm{Rb}$ $4{D}_{5/2}$
  state,} Opt. Lett. \textbf{32}, 2810--2812 (2007).

\bibitem{Wang_2014}
J.~Wang, H.~Liu, G.~Yang, B.~Yang, and J.~Wang, \enquote{Determination of the
  hyperfine structure constants of the $^{87}\mathrm{Rb}$ and
  $^{85}\mathrm{Rb}$ $4{D}_{5/2}$ state and the isotope hyperfine anomaly,}
  Phys. Rev. A \textbf{90}, 052505 (2014).

\bibitem{Lee_2015}
W.-K. Lee and H.~S. Moon, \enquote{Measurement of absolute frequencies and
  hyperfine structure constants of $4{D}_{5/2}$ and $4{D}_{3/2}$ levels of
  $^{87}\mathrm{Rb}$ and $^{85}\mathrm{Rb}$ using an optical frequency comb,}
  Phys. Rev. A \textbf{92}, 012501 (2015).

\bibitem{Roy_2015}
R.~Roy, \enquote{An integrated atom chip for the detection and manipulation of
  cold atoms using a two-photon transition,} Ph.D. thesis, Centre for Quantum
  Technologies, National University of Singapore (2015).

\bibitem{Srivathsan_2013}
B.~Srivathsan, G.~K. Gulati, B.~Chng, G.~Maslennikov, D.~Matsukevich, and
  C.~Kurtsiefer, \enquote{Narrow band source of transform-limited photon pairs
  via four-wave mixing in a cold atomic ensemble,} Phys. Rev. Lett.
  \textbf{111}, 123602 (2013).

\bibitem{GulatiPRA_2014}
G.~K. Gulati, B.~Srivathsan, B.~Chng, A.~Cer\`e, D.~Matsukevich, and
  C.~Kurtsiefer, \enquote{Generation of an exponentially rising single-photon
  field from parametric conversion in atoms,} Phys. Rev. A \textbf{90}, 033819
  (2014).

\end{thebibliography}
\end{document}